\documentclass[reprint,aps,ams,amsmath,prx,twocolumn,longbibliography,superscriptaddress]{revtex4-1}
\usepackage{graphics}
\usepackage{graphicx}
\usepackage{epsfig}
\usepackage{amsmath}
\usepackage{relsize}
\usepackage{amssymb}
\usepackage{amsfonts}
\usepackage{bm}
\usepackage{color}
\usepackage{bbm}
\usepackage{hyperref}
\usepackage[normalem]{ulem}
\usepackage{comment}
\usepackage{soul}
\def\thefootnote{*}\footnotetext{J.H. and K.N.N.  contributed equally to this work.}

\begin{document}

\title{Anomalous Josephson effect in planar noncentrosymmetric superconducting devices}

\author{Jaglul Hasan\thefootnote{}}
\affiliation{Department of Physics, University of Wisconsin-Madison, Madison, Wisconsin 53706, USA}

\author{Konstantin N. Nesterov\thefootnote{}}
\affiliation{Department of Physics, University of Wisconsin-Madison, Madison, Wisconsin 53706, USA}
\affiliation{Bleximo Corporation, Berkeley, California 94710, USA}

\author{Songci Li}
\affiliation{Department of Physics, University of Wisconsin-Madison, Madison, Wisconsin 53706, USA}

\author{Manuel Houzet}
\affiliation{Universit\'{e} Grenoble Alpes, CEA, Grenoble INP, IRIG, Pheliqs, 38000 Grenoble, France}

\author{Julia S. Meyer}
\affiliation{Universit\'{e} Grenoble Alpes, CEA, Grenoble INP, IRIG, Pheliqs, 38000 Grenoble, France}

\author{Alex Levchenko}
\affiliation{Department of Physics, University of Wisconsin-Madison, Madison, Wisconsin 53706, USA}

\date{November 19, 2022}

\begin{abstract}
In two-dimensional electron systems with broken inversion and time-reversal symmetries, a Josephson junction reveals an anomalous response: the supercurrent is nonzero even at zero phase difference between two superconductors. We consider details of this peculiar phenomenon in the planar double-barrier configurations of hybrid circuits, where the noncentrosymmetric normal region is described in terms of the paradigmatic Rashba model of spin-orbit coupling. We analyze this anomalous Josephson effect by means of both the Ginzburg-Landau formalism and the microscopic Green's functions approach in the clean limit. The magnitude of the critical current is calculated for an arbitrary in-plane magnetic field orientation, and anomalous phase shifts in the Josephson current-phase relation are determined in terms of the parameters of the model in several limiting cases.   \\

\textit{Spin-Coherent Phenomena in Semiconductors: Special Issue in Honor of Emmanuel I. Rashba}. 
\end{abstract}

\maketitle

%########################################################################
%########################################################################
%########################################################################
%########################################################################
%########################################################################

\section{Introduction}

In Josephson junctions (JJ) of conventional $s$-wave superconductors, the supercurrent-phase relation $j(\phi)$ is expected to obey rather general properties that depend neither on the junction’s geometry nor on the scattering processes taking place in the junction region, in other words, they apply to tunnel, ballistic, and diffusive junctions \cite{GKI2004}. (i) The first basic property follows from the $2\pi$ periodicity of the superconducting order parameter, which implies that   $j(\phi)=j(\phi+2\pi)$. (ii) The second property reflects the fact that changing the direction of the phase gradient applied across the junction reverses the direction of the superflow, $j(\phi)=-j(-\phi)$, and therefore the supercurrent-phase relation is an odd function. (iii) The current must vanish at all integer phases modulo $2\pi$, namely, $j(2\pi n)=0$ for $n\in\mathbb{Z}$. This condition states an obvious thermodynamic requirement that a finite supercurrent is induced only by a nonzero phase gradient, so it vanishes for $\phi=0$, and then by virtue of periodicity must vanish at other phases multiple of $2\pi$. (iv) The combination of the first two properties dictates that $j(\pi n)=0$ for $n\in\mathbb{Z}$; therefore it is sufficient to consider $j(\phi)$ only in the interval $0<\phi<\pi$. Additionally, it should be noted that in general, symmetries of the full Hamiltonian describing a Josephson junction, or their absence, can be related to the particular features in the pattern of the supercurrent-phase relation. 

The anomalous Josephson effect (AJE), where the above-formulated properties of the supercurrent-phase relation are altered, can be realized in superconductors with broken time-reversal symmetry, leading to spontaneous currents. There are two kinds of systems where these effects have been discussed: (i) JJs between magnetic superconductors
with unconventional pairing symmetry \cite{Geshkenbein1986,Yip1995,Sigrist1998,Kashiwaya2000}; (ii) superconductor-ferromagnet-superconductor (SFS) junctions, and their more complex hybrids with additional noncollinear ferromagnetic layers and insulating barriers \cite{Buzdin1982,Ryazanov2001,Buzdin2005,Braude2007,Houzet2007,Birge2016}. In particular, in the original work of Geshkenbein and Larkin \cite{Geshkenbein1986} devoted to JJs based on heavy-fermion superconductors, the following current-phase relation was predicted:
\begin{equation}\label{eq:i-phi-GL}
j(\phi)=j_1\sin\phi+j_2\cos\phi=j_c\sin(\phi+\phi_0),    
\end{equation}
where $j_c=\sqrt{j^2_1+j^2_2}$ is the critical current, and $\phi_0=\arctan(j_2/j_1)$ is the anomalous phase shift, whose microscopic form depends on the system under consideration and specific model assumptions. In general, the current-phase relation is not simply sinusoidal. Indeed, the contribution of higher-order harmonics may be non-negligible, which is often the case at temperatures much lower than the critical. Therefore the generalized form of Eq. \eqref{eq:i-phi-GL} can be presented as the Fourier series,
\begin{equation}\label{eq:i-phi-fourier}
j(\phi)=\sum_{n\geq 1} \left[j_{1n}\sin(n\phi)+j_{2n}\cos(n\phi)\right], 
\end{equation}
and contributions with $j_{2n}$ are typically present as long as time-reversal symmetry is broken. 

A different kind of AJE was proposed later in Refs. \cite{Buzdin2008,Reynoso2008}; see also important preceding works \cite{Krive2004,Krive2005}. 
The key insight of those works is that current-phase relation of the type Eq. \eqref{eq:i-phi-fourier} can be realized even in junctions of conventional superconductors when the normal layer between them is a noncentrosymmetric metal, i.e., with broken inversion symmetry. As a guiding example, calculations were presented for a weak link with Rashba-type spin-orbit coupling \cite{Rashba1984}, and Eq. \eqref{eq:i-phi-GL} was derived microscopically in the quasi-one-dimensional geometry. To separate this anomalous Josephson effect from that in unconventional JJs, the term ``$\phi_0$-junction" was introduced \cite{Buzdin2008}.  

In recent years we received compelling experimental verification of these anomalous Josephson phenomena in various heterojunctions \cite{Szombati2016,Aubin2019,Shabani2020,Braggio2020,Giazotto2020,Dvir2021,Kim2021,Checkley2021,Strunk2022}. These devices represent a diverse class of systems that differ from each other by material components, dimensionality, quality of contacts, and purity of interlayers between superconducting banks, thus reflecting the prevalence and robustness of the aforementioned effects. Theoretical studies address a broad spectrum of questions related but not limited to (i) types of spin-orbit interaction, including spin-active interfaces; (ii) effects of impurities; and (iii) electronic band structure, in particular topological properties. There are a number of notable theoretical contributions to this topic, and we can highlight studies of the AJE in quantum dots \cite{Zazunov2009,Brunetti2013}, semiconducting nanowires \cite{Yokoyama2013,Ojanen2013,Yokoyama2014,Campagnano2015,Mironov2015,Wu2016,Nesterov2016}, and topological \cite{Tanaka2009,Linder2010,BlackSchaffer2011,Nussbaum2014,Lu2015,Dolcini2015,Marra2016,Zyuzin2016} and nontopological systems \cite{Bezuglyi2002,Dimitrova2006,Galaktionov2008,Brydon2008,Grein2009,Konschelle2009,Liu2010a,Liu2010b,Margaris2010,Liu2011,Reynoso2012,Bergeret2015,Konschelle2015,Rasmussen2016,Silaev2017} that involve a combination of unconventional superconductors, topological surface or edge states, and ferromagnets.   

The continuous improvement in the quality of materials, where the electron mean free path is comparable to or even may exceed the dimensions of the junction, call for the investigation of AJE in the clean limit, which thus far has received very limited theoretical attention. This task is accomplished in the present work, and the rest of the paper is organized as follows. In Sec. \ref{sec:GL} we apply Ginzburg-Landau (GL) phenomenology to address the anomalous Josephson effect in the two-dimensional electron gas (2DEG) with Rashba-type spin-orbit coupling. Even though the GL formalism has its limitations, it gives us an opportunity to fully analytically investigate the field dependence of the critical current and the phase shift in the two-dimensional geometry. The principal results of this work are presented in Sec. \ref{sec:microscopics}, where we develop a microscopic theory of the AJE based on the Gor'kov equations for two complementary junction geometries. This analysis expands previous considerations of the AJE that exploited semiclassical approximations for ballistic (Eilenberger limit) and diffusive (Usadel limit) systems. In Sec. \ref{sec:summary} we provide summary of our findings in comparison to earlier related works.     

%$$$$$$$$$$$$$$$$$$$$$$$$$$$$$$$$$$$$$$$$$$$$$$$$$$$$$$$$$$$$$$$$$$$$
\begin{figure}[t!]%[hb]
\includegraphics[width=0.48\textwidth]{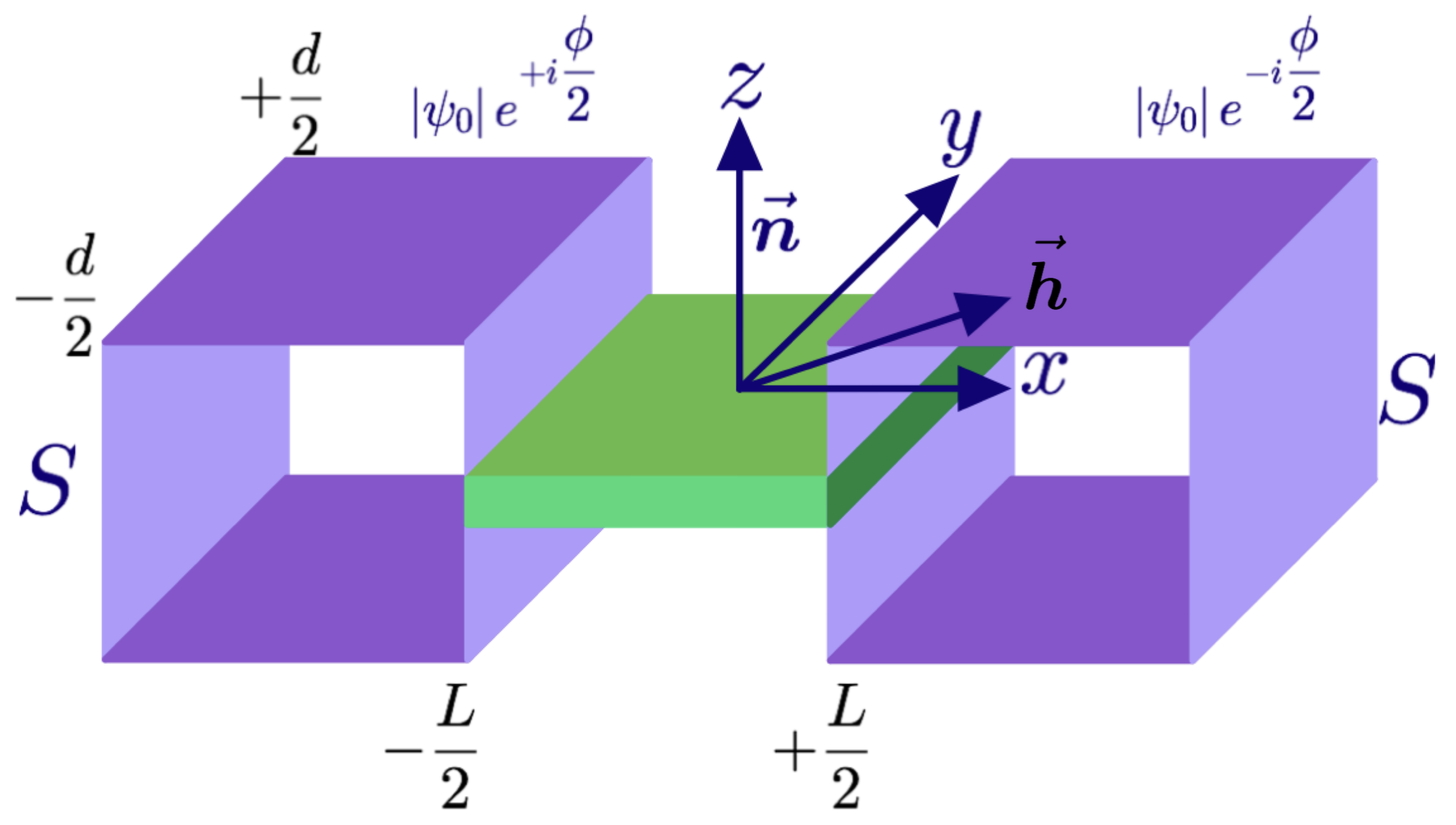} 
\caption{Geometry of a planar SINIS Josephson junction with Rashba-type 2DEG as the normal region. The in-plane magnetic field ${\boldsymbol{h}}$ is directed arbitrarily in the $x-y$ plane and the vector $\boldsymbol{n}$ is along the $z$ axis. The total length of the normal region is $L$ and its total width is $d$. The complex superconducting order parameter in the leads is $|\psi_0|e^{\pm i\phi/2}$ so that $\phi$ is the total phase difference across the junction.}  \label{fig1}
\end{figure}
%$$$$$$$$$$$$$$$$$$$$$$$$$$$$$$$$$$$$$$$$$$$$$$$$$$$$$$$$$$$$$$$$$$$$

%########################################################################
%########################################################################
%########################################################################
%########################################################################
%########################################################################

\section{Ginzburg-Landau formalism} \label{sec:GL}

To elucidate the unusual properties of the AJE, we start with the simple Ginzburg-Landau (GL) model before delving into the microscopic calculation. The geometry that we consider is depicted schematically in Fig. \ref{fig1} in which a normal region of two-dimensional electron gas (2DEG) of length $L$ and width $d$ is flanked by two conventional $s$-wave superconducting banks. Let us consider the situation where the time reversal symmetry (TRS) in the system is broken by an in-plane magnetic field $\boldsymbol{h}$ and the space inversion symmetry in the normal region is broken by the presence of a Rashba-type spin-orbit coupling (SOC) term \cite{Rashba1984} given by $\alpha[\boldsymbol{\sigma} \times \boldsymbol{p}] \cdot \boldsymbol{n}$. Here $\boldsymbol{\sigma}$ is the Pauli spin matrix-vector, $\boldsymbol{p}$ is the particle momentum, $\boldsymbol{n}$ is the unit vector along the direction of the asymmetric potential gradient, and the parameter $\alpha$ denotes the strength of the spin-orbit interaction, which has units of velocity. 

In the presence of SOC, the GL free energy $\Omega$ was derived by Edelstein \cite{Edelstein1996} (see also related works Refs. \cite{Samokhin2004,Dimitrova2007,Meyer2015,Edelstein2021}):
\begin{equation}\label{eq:GL1}
\begin{split}
\Omega\left(\psi, \psi^{*}\right)=\int d\bm{r}\Biggl[a|\psi|^{2}+\frac{b}{2}|\psi|^{4}+\frac{1}{4m}|{\bm{\partial}} \psi|^{2} \\
-\epsilon \left[\boldsymbol{n}\times\boldsymbol{h}\right] \cdot\left\{\psi(\bm{\partial} \psi)^{*}+\psi^{*}(\bm{\partial}\psi)\right\}+\frac{\boldsymbol{h}^{2}}{8 \pi}\Biggr],
\end{split}
\end{equation}
where $\psi(\bm{r})$ is the spatially inhomogeneous superconducting order parameter, $\boldsymbol{h}=\boldsymbol{\nabla} \times \boldsymbol{A}$ where $\boldsymbol{A}$ is the vector potential, and $\bm{\partial}=$ $-i\bm{\nabla}-2e\bm{A}$ is the gauge invariant derivative. Here and in what follows, we work in the units $\hbar=k_B=c=1$. This form of the GL functional applies to both clean and disordered superconductors. The difference is in the dependence of the expansion coefficients $a,b,\epsilon$, and also the gradient term, on the strength of spin-orbit $\alpha$, critical temperature $T_c$ of a superconductor, and elastic scattering time $\tau$ induced by disorder potential. The conventional part of the GL functional, namely, the first three terms in Eq. \eqref{eq:GL1}, weakly depends on the SOC. In contrast, the coefficient $\epsilon=\frac{\alpha}{v_{F}p_F} f_d(\frac{\alpha p_{F}}{T_{c}},T_c\tau)$, with $v_{F}$ and $p_{F}$ being the Fermi velocity and the Fermi momentum, respectively, depends sensitively on $\alpha$. The asymptotic form of the function $f_d$ is established for two- and three-dimensional superconductors in various limiting cases, see Refs. \cite{Edelstein1996,Meyer2015,Edelstein2021}
for details. Both parameters $\alpha p_{F} / T_{c}$ and $T_c\tau$ can be of the order of unity in materials.  

The free energy functional in Eq. (\ref{eq:GL1}) must be minimized with respect to the order parameter $\psi^{*}$ and the vector potential $\boldsymbol{A}$ to get the equilibrium GL equations. Therefore, varying \eqref{eq:GL1} with respect to $\psi^{*}$ and $\boldsymbol{A}$ and setting that variation equal to zero for arbitrary variations $\delta \psi^{*}$ and $\delta \boldsymbol{A}$, the two GL equations are obtained in the form 
\begin{equation}\label{eq:GL2}
\begin{split}
\frac{1}{4m}\bm{\partial}^{2} \psi(\boldsymbol{r})-\frac{1}{2m}\boldsymbol{Q} \cdot \bm{\partial} \psi(\boldsymbol{r})+a \psi(\boldsymbol{r})\\
+b|\psi(\boldsymbol{r})|^{2} \psi(\boldsymbol{r})=0,
\end{split}
\end{equation}
\begin{equation}\label{eq:GL3}
\begin{split}
\boldsymbol{j}=\frac{e}{2m}\left\{\psi(\bm{\partial} \psi)^{*}+\psi^{*}(\bm{\partial}\psi)\right\}-4 e \epsilon|\psi|^{2}(\boldsymbol{n} \times \boldsymbol{h}) \\ 
-\epsilon  \operatorname{\boldsymbol{curl}}[\boldsymbol{n} \times \{\psi(\bm{\partial}\psi)^{*}+\psi^{*}(\bm{\partial}\psi)\}],
\end{split}
\end{equation}
with the boundary condition,
\begin{equation}\label{eq:GL4}
\left[(\bm{\partial}-\boldsymbol{Q})\psi\right] \cdot \boldsymbol{n}_{b}=0.
\end{equation}
The unit vector $\boldsymbol{n}_{b}$ is the normal vector at the system boundary. The wave vector $\boldsymbol{Q}=4m\epsilon\left[\boldsymbol{n} \times \boldsymbol{h}\right]$ represents an emergent scale for this GL theory with Rashba SOC and Zeeman field. Its presence induces a spatially modulated helical superconducting phase given by $\psi \propto e^{i \boldsymbol{Q} \cdot \boldsymbol{r}}$. We recall that the boundary conditions [Eq. (\ref{eq:GL4})] on these equations are obtained from the condition that the surface integrals in the variation $\delta \Omega$ are zero. As a result of this condition, the normal component of the supercurrent density (\ref{eq:GL3}) at the boundary of the superconductor with vacuum is $\boldsymbol{j}\cdot \boldsymbol{n}_{b}=0$.  

This framework was used in the original work \cite{Buzdin2008} to derive the anomalous Josephson current in a quasi-one-dimensional geometry with rigid boundary conditions. Below we extend these results to a full two-dimensional geometry with an arbitrarily oriented field in the plane of the 2DEG and also with more general boundary conditions: 
\begin{equation}\label{eq:GL5}
\left[(\bm{\partial}-\boldsymbol{Q}) \psi\right] \cdot \boldsymbol{n}_{b}=\frac{1}{il_t}\psi. 
\end{equation} 
Here $l_t$ is the extrapolation length to the point outside the boundary at which the order parameter $\psi$ would vanish if it maintained the slope it had at the surface \cite{Tinkham}. The value of $l_t$ depends on the nature of the material to which the interface is made, approaching zero for a magnetic material or in the case of a high density of defects in the interface (Dirichlet boundary condition), and infinity for an insulator or vacuum (Neumann boundary condition), with normal metals lying in between.

%$$$$$$$$$$$$$$$$$$$$$$$$$$$$$$$$$$$$$$$$$$$$$$$$$$$$$$$$$$$$$$$$$$$$
\begin{figure}[t]
\includegraphics[width=0.5\textwidth]{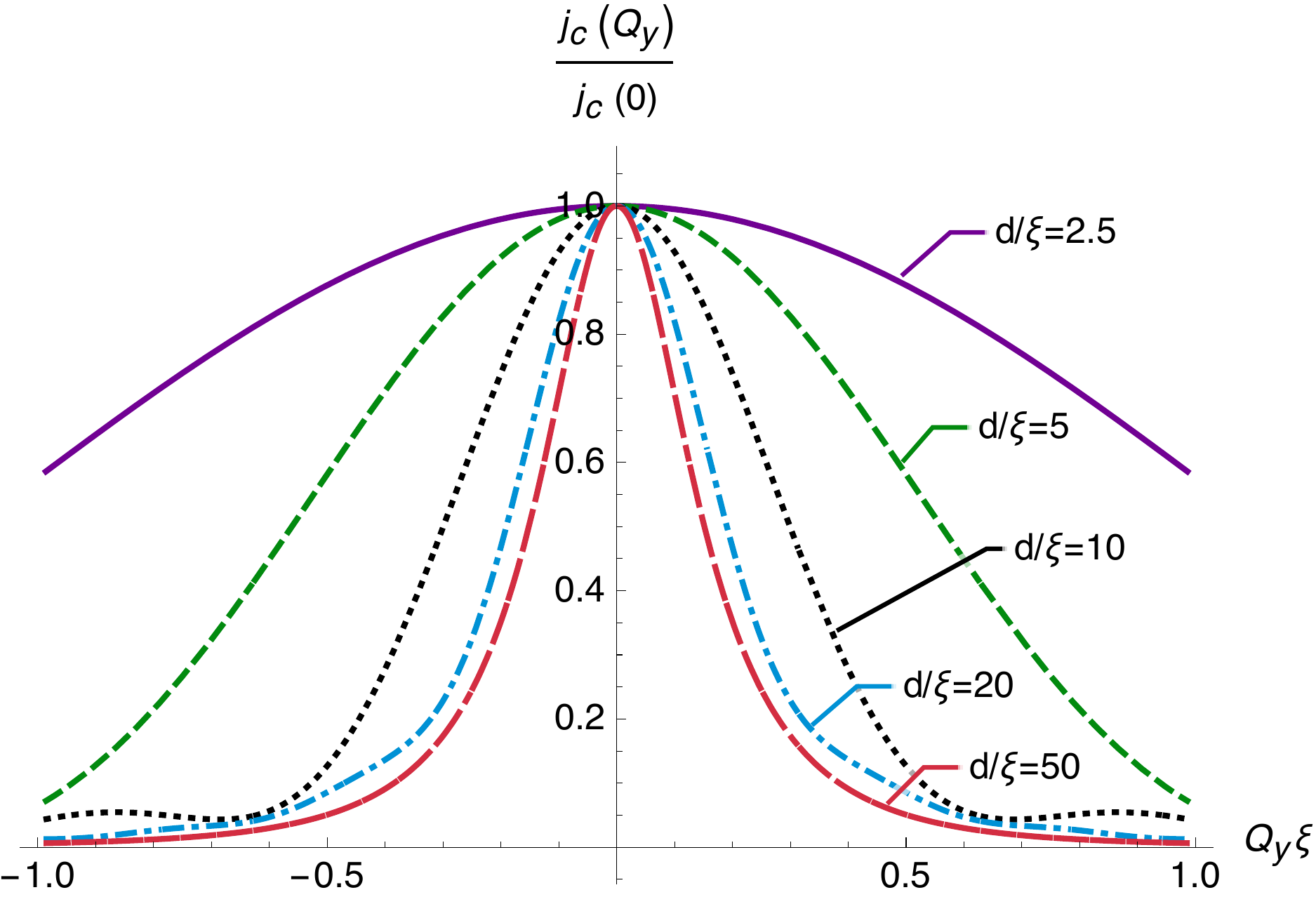} 
\caption{
Normalized critical current density, $j_c(Q_y)/j_c(0)$, is plotted as a function of the Zeeman field, $Q_y\propto h_x$, for several different values of the sample width $d$, which is measured in the units of the superconducting coherence length $\xi$ and for a fixed ratio $L/\xi=5$. 
}  \label{fig2}
\end{figure}
%$$$$$$$$$$$$$$$$$$$$$$$$$$$$$$$$$$$$$$$$$$$$$$$$$$$$$$$$$$$$$$$$$$$$

To calculate the Josephson current in the geometry of Fig. \ref{fig1}, we can neglect the orbital effect of the field. We can also neglect the nonlinear term proportional to $b$ in Eq. \eqref{eq:GL2}. The solution for $\psi(x,y)$ can be written as the series 
\begin{equation}\label{eq:GL6}
\psi=e^{i\bm{Qr}}
\sum_n \left(A_{n}e^{x/\xi_{n}}+B_{n}e^{-x/\xi_{n}}\right)\cos[q_n(y+d/2)]
\end{equation}
where 
\begin{equation}\label{eq:GL7}
q_n=\pi n/d,\qquad \xi^{-1}_n=\sqrt{4 m a -Q^{2}+q^2_n},
\end{equation}
with $Q=\sqrt{Q_x^2+Q_y^2}=4m\epsilon \sqrt{h^2_x+h^2_y}=4m\epsilon h$. The expansion coefficients $A_n$ and $B_n$ are easily calculated by using appropriate boundary conditions from Eq. \eqref{eq:GL5} for the two SN interfaces,
\begin{equation}\label{eq:GL8}
\begin{gathered}
\left.\left(\frac{\partial \psi}{\partial x}-i Q_{x} \psi\right)\right|_{x=\pm \frac{L}{2}}=\pm \frac{1}{l_t}\left|\psi_{0}\right| e^{\mp i \frac{\phi}{2}}, \\
\end{gathered}
\end{equation}
where, in the superconducting banks to the left and right sides of the normal region, $\psi(x<-\frac{L}{2},y)=\left|\psi_{0}\right| e^{+i \frac{\phi}{2}}$ and $\psi(x>\frac{L}{2},y)=\left|\psi_{0}\right| e^{-i \frac{\phi}{2}}$. In these solution we require $4ma>Q^{2}$, because the smallest value of $|n|=0$. This is equivalent to the condition $Q\xi<1$, which means that the length scale characterized by the inverse of the wave vector $Q$ must be greater than the superconducting coherence length $\xi=\sqrt{\frac{1}{4ma}}$. Equation (\ref{eq:GL3}) can now be used to find the current density:
\begin{equation}\label{eq:GL9}
j_x(y)=\frac{e}{2mi}\left(\psi^* \frac{\partial\psi}{\partial x} -\psi \frac{\partial\psi^{*}}{\partial x} \right)-\frac {e Q_x}{m}|\psi|^{2}.
\end{equation}
Since the current density is inhomogeneous, we are interested in a current density averaged over the sample width $d$:
\begin{equation}\label{eq:GL10}
\begin{gathered}
j(\phi)=\frac{1}{d} \int\limits_{-d/2}^{+d/2} j_{x}(y) dy=\frac{e}{2mi}\sum_{n \in \mathbb{Z}}\left[\frac{2}{\xi_{n}}\left(B_{n}^{*} A_{n}-A_{n}^{*} B_{n}\right)\right].
\end{gathered}
\end{equation}
This can be simplified to get the form of the anomalous Josephson effect in the $\phi_0$ junction
\begin{equation}\label{eq:GL11}
j(\phi)=j_c(Q_y)\sin(\phi+\phi_0).
\end{equation}
In this model, the critical current density $j_c$ and the anomalous phase shift $\phi_0$ are given by
\begin{equation}\label{eq:GL12}
\begin{split}
&\frac{{j}_{c}(Q_y)}{j_c(0)}=\frac{\sin^2(Q_yd/2)}{(Q_yd/2)^2} \\ 
&+\sum_{n\geq 1}
\frac{4(Q_yd)^2\left[1-(-1)^{n} \cos \left(Q_y d\right)\right]}{\left(Q_y^{2} d^{2}-n^{2} \pi^{2}\right)^{2}}\frac{\xi_n\sinh(L/\xi_0)}{\xi_0\sinh(L/\xi_n)}
\end{split}
\end{equation}
and
\begin{equation}\label{eq:GL13}
\phi_0=Q_x L=4m\epsilon h_y L,
\end{equation}
where $\xi_n$ is given by Eq. (\ref{eq:GL7}). From Eqs. (\ref{eq:GL12}) and (\ref{eq:GL13}) we note that the Zeeman field $\boldsymbol{h}$ has dual effect: parallel to the SN boundary component, $Q_{x}\propto h_y$, defines the anomalous phase shift $\phi_0$, while the perpendicular to the SN interface component, $Q_{y}\propto h_x$, governs the current density $j_c$ modulation, see Fig. \ref{fig2} for the illustration. These results can be further generalized to superconducting leads with Rashba coupling and Zeeman field. In that case, the vector $\bm{Q}$ in the final expressions should be replaced by the difference between the corresponding vectors $\bm{Q}$ in the superconducting and normal parts.

The anomalous phase shift matches with the earlier result of Ref. \cite{Buzdin2008}. The critical current can be also recovered if we single out the $n=0$ contribution from the sum over $n \in \mathbb{Z}$ in Eq. (\ref{eq:GL12}), take the $d \to 0$ limit and consider the rigid boundary conditions, i.e., $\psi(x=-\frac{L}{2},y)=\left|\psi_{0}\right| e^{+i \frac{\phi}{2}}$ and $\psi(x=\frac{L}{2},y)=\left|\psi_{0}\right| e^{-i \frac{\phi}{2}}$. 

%########################################################################
%########################################################################
%########################################################################
%########################################################################
%########################################################################

\section{Microscopic approach}\label{sec:microscopics}

In the context of various possible Josephson microconstrictions, the example considered in the previous section corresponds to the case of SINIS junctions, where ``I" denotes an insulating tunnel barrier. The microscopic calculations of supercurrent-phase relations in such devices were originally carried out in several important works. In Ref. \cite{ALO1969} a tunneling Hamiltonian was used and the critical current was calculated for the diffusive limit of transport. In Ref. \cite{KL1988} boundary conditions were derived in the semiclassical limit, and applications to the Josephson effect were given based on the solution of Usadel equations. In Ref. \cite{BG2000}, the general solution for ballistic electronic transport through double-barrier junctions was elaborated in the phase-coherent limit (see also review Ref. \cite{GKI2004} and additional references therein that expand on the topic). Following these works, we consider below planar SINIS junctions in a particular model of extended tunnel barriers with the focus on the anomalous phase shifts.    

%@@@@@@@@@@@@@
\subsection{Model Hamiltonian}
%@@@@@@@@@@@@@

The total Hamiltonian for the SINIS system under consideration consists of three main parts: 
\begin{equation}\label{eq:IIIA1}
\hat{H}=\hat{H}_{S}+\hat{H}_N+\hat{V}_{T}.
\end{equation}
Here $\hat{H}_{S}=\hat{H}_{L}+\hat{H}_{R}$ represents left-$(L)$ and right-$(R)$ conventional $s$-wave superconducting electrodes described within the BCS model with equal pairing gaps $\Delta_{L}=\left|\Delta\right| e^{i \phi_{L}}$ and $\Delta_{R}=\left|\Delta\right| e^{i \phi_{R}}$, and overall phase difference $\phi=\phi_L-\phi_R$ \cite{Tinkham}. 

The Hamiltonian of the normal layer, $\hat{H}_N$, describes a two-dimensional metal in the presence of a Zeeman field and with a Rashba term:
\begin{equation}\label{eq:IIIA15}
\hat{H}_N=\sum_{\sigma \sigma^{\prime}} \int d\bm{r} \Psi_{N\sigma}^{\dagger}(\bm{r}) \hat{h}_{\sigma \sigma^{\prime}} \Psi_{N\sigma^{\prime}}(\bm{r}),
\end{equation}
with 
\begin{equation}\label{eq:IIIA16}
\hat{h}_{\sigma \sigma^{\prime}}=\frac{\hat{\bm{p}}^{2}}{2 m} \delta_{\sigma \sigma^{\prime}}+\alpha[\hat{\bm{p}} \times \bm{n}] \cdot \bm{\sigma}+\bm{h} \cdot \bm{\sigma}. 
\end{equation}
This is exactly the same model as used in Sec. \ref{sec:GL} that led to an effective GL-free energy in Eq. \eqref{eq:GL1}. In accordance with the usual convention, $\Psi_{N\sigma}^{\dagger}$ and $\Psi_{N\sigma}$ in Eq. \eqref{eq:IIIA15} describe electron creation and annihilation field operators, and $\sigma$ is the spin index. For the geometry of Fig. \ref{fig1} with the $z$ axis normal to the plane of the metal, the unit vector $\bm{n}$ has only a $z$ component. Therefore if we consider a Zeeman field in the $xy$ plane, $\bm{h}=(h \cos \varphi, h \sin \varphi, 0)$, the single-particle operator $\hat{h}$ in Eq. \eqref{eq:IIIA16} can be written as the following matrix: 
\begin{equation}\label{eq:IIIA18}
\hat{h}=\left(\begin{array}{cc}
\frac{\hat{p}_{x}^{2}}{2 m}+\frac{\hat{p}_{y}^{2}}{2 m} & \alpha \hat{p}_{y}+i \alpha \hat{p}_{x}+h e^{-i \varphi} \\
\alpha \hat{p}_{y}-i \alpha \hat{p}_{x}+h e^{i \varphi} & \frac{\hat{p}_{x}^{2}}{2 m}+\frac{\hat{p}_{y}^{2}}{2 m}
\end{array}\right).
\end{equation}

The coupling between the electrodes and the normal region is described by the tunneling part of the Hamiltonian. For the model of an extended tunnel junction that is translationally invariant along $y$ we have 
\begin{equation}
\hat{V}_{T}=\hat{X}_{L N}+\hat{X}_{N R}+\hat{X}_{L N}^{\dagger}+\hat{X}_{N R}^{\dagger}
\end{equation}\label{eq:IIIA2}
with \cite{PradaSols2004}
\begin{subequations}
\begin{align}
\hat{X}_{L N}=t \sum_{\sigma} \int_{b 1} d\bm{r}_{b 1} \frac{\partial}{\partial \bm{r}_{n}} \Psi_{L \sigma}^{\dagger}\left(\bm{r}_{b 1}\right) \frac{\partial}{\partial \bm{r}_{n}} \Psi_{N \sigma}\left(\bm{r}_{b 1}\right) \label{eq:IIIA3}, \\ 
\hat{X}_{N R}=t \sum_{\sigma} \int_{b 2} d\bm{r}_{b 2} \frac{\partial}{\partial \bm{r}_{n}} \Psi_{N \sigma}^{\dagger}\left(\bm{r}_{b2}\right) \frac{\partial}{\partial \bm{r}_{n}} \Psi_{R \sigma}\left(\bm{r}_{b 2}\right).\label{eq:IIIA4}
\end{align}
\end{subequations}
The integrals are taken along the junction interfaces, and $\bm{r}_{n}$ is normal to the boundary. The normal derivatives have to be understood as being taken from the side where the corresponding function is defined. In momentum space, tunnel matrix elements have the form $t_{kk'}\propto t k_xk'_x\delta(k_y-k'_y)$. In the normal state, the tunnel conductance is given by $g=\frac{32}{3\pi^2}m^2e^2t^2p^3_FS_d$, where $S_d$ is the total area of the interface. A similar model without the normal derivatives would imply longitudinal momentum independent tunneling matrix elements. The rationale for the extended model is discussed extensively in Ref. \cite{PradaSols2004}. 

In the calculations below, we work in perturbation theory with respect to the Zeeman field and strength of spin-orbit interaction as compared to the Fermi energy, namely, $\{h,\alpha p_F\}\ll \varepsilon_F$. In particular, this enables us to neglect the effect of the field on the suppression of the order parameter in the leads. For the hierarchy of relevant length scales we explore various relations between the thermal length $l_T=v_F/T$, the superconducting coherence length $\xi$, the spin-orbit length $l_{so}=1/m\alpha$, and the distance $L$ between the tunnel contacts.    

%@@@@@@@@@@@@@@
\subsection{Josephson current}
%@@@@@@@@@@@@@@

The operator of the current flowing from $L$ to $R$ is given by a commutator through the equation of motion,
\begin{equation}\label{eq:IIIA5}
\hat{I}_{L \rightarrow R}=e\hat{\dot{N}}_{L}=ie\left[\hat{V}_{T}, \hat{N}_{L}\right],
\end{equation}
where $\hat{N}_{L}$ is the operator of the electron number in the left lead. Therefore, $\hat{I}_{L \rightarrow R}=-ie\left(\hat{X}_{L N}-\hat{X}_{L N}^{\dagger}\right)$, and consequently, the thermal average for the expectation value of the tunneling current can be written as follows:
\begin{equation}\label{eq:IIIA7}
\left\langle\hat{I}_{L \rightarrow R}\right\rangle=I_{L \rightarrow R}=2e\operatorname{Im}\langle\hat{X}_{L N}\rangle.
\end{equation}
To calculate this average, we use the interaction picture representation \cite{ALO1969} 
\begin{equation}\label{eq:IIIA8}
I_{L \rightarrow R}=2 e \operatorname{Im} \frac{\operatorname{Tr}\left[\hat{X}_{L N} e^{-\beta(\hat{H}_0-\mu \hat{N})} \hat{\mathcal{U}}_I(\beta)\right]}{\operatorname{Tr}\left\{e^{-\beta(\hat{H}_0-\mu \hat{N})} \hat{\mathcal{U}}_I(\beta)\right\}},
\end{equation}
where $\hat{H}_0$ denotes the system Hamiltonian without the tunneling part, and $\hat{\mathcal{U}}_I(\beta)=\operatorname{T_{\tau} exp}\left[-\int_{0}^{\beta} \hat{V}_{T}(\tau) d \tau\right]$, where $\operatorname{T_{\tau}}$ denotes time-ordering in imaginary time. 

\begin{figure}[t!]
\includegraphics[width=0.48\textwidth]{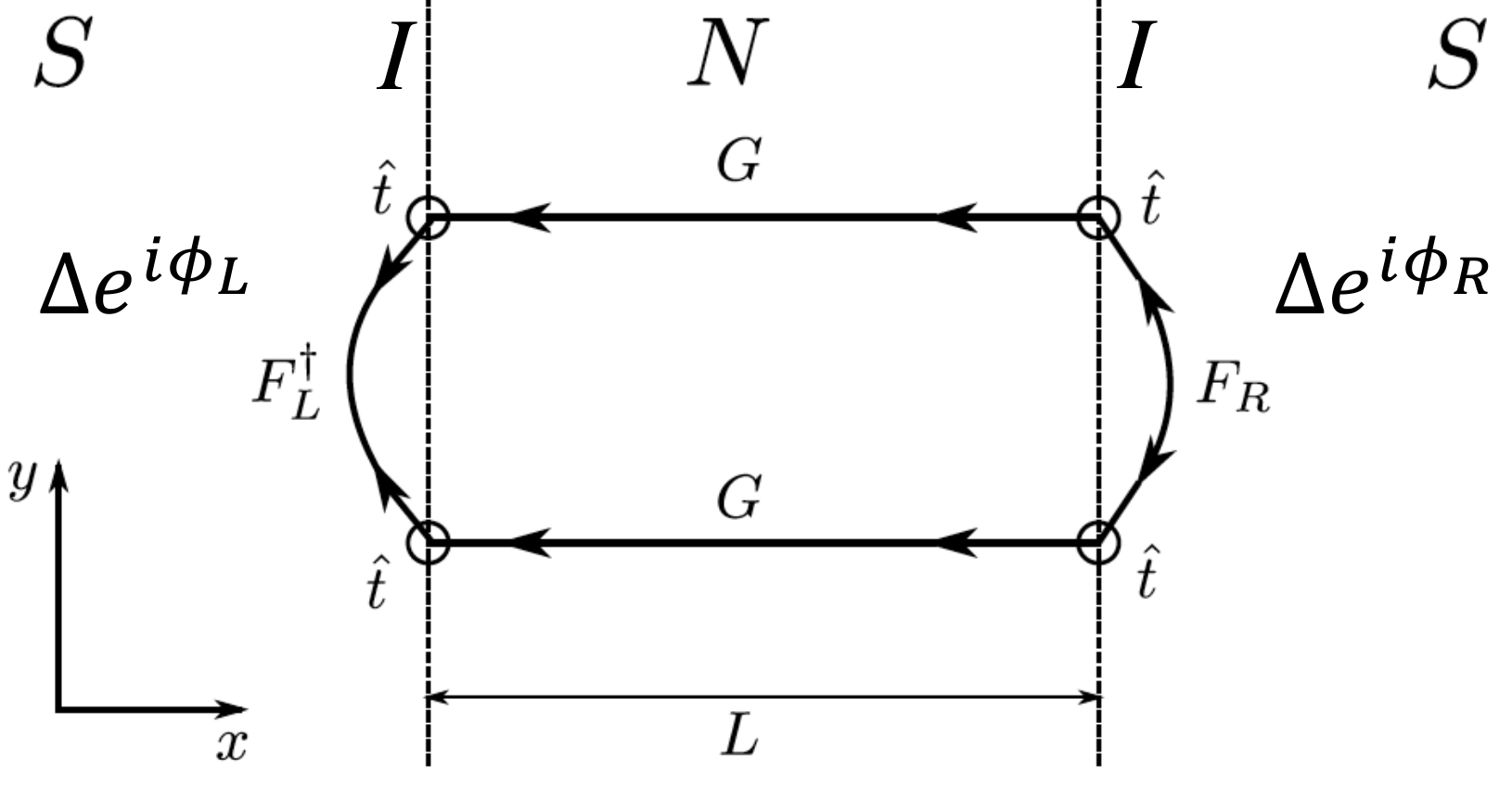} 
\caption{Diagram representing the leading order in tunneling contribution to the Josephson current $j(\phi)$ of an SINIS junction. Superconducting leads marked by $S$ are assumed to have equal gaps $\Delta$ but different phases $\phi=\phi_{L}-\phi_R$. Tunneling events at  the insulating interfaces $I$ are marked by $\hat{t}$. $G$ and $F$ represent normal and anomalous Green's functions.}  \label{fig3}
\end{figure}

In the expression of Eq. \eqref{eq:IIIA8}, the first nonvanishing contribution to the current appears in fourth order in the tunneling matrix element $t$.  We thus expand the exponential in powers of $\hat{V}_T$ and determine that the following combinations of $\hat{X}$'s and $X^{\dagger}$'s contribute to the current: 
\begin{equation}\label{eq:IIIA10}
\begin{split}
&I_{L \rightarrow R}
\equiv I=-e \operatorname{Im}\biggl[\int_{0}^{\beta} d \tau_{1} d \tau_{2} d \tau_{3}\\
&\left\langle \operatorname{T_{\tau}} \hat{X}_{N R}\left(\tau_{1}\right) \hat{X}_{N R}\left(\tau_{2}\right) \hat{X}_{L N}\left(\tau_{3}\right) \hat{X}_{L N}(0)\right\rangle\biggr].
\end{split}
\end{equation}
At this point we apply Wick's theorem to contract field operators and express them in terms of the normal and anomalous Gor'kov Green's functions \cite{Mahan}: $G_{ \sigma \sigma^{\prime}}\left(\bm{r}, \bm{r}^{\prime}, \tau-\tau^{\prime}\right)=-\left\langle \operatorname{T_{\tau}}\Psi_{ \sigma}\left(\bm{r}, \tau\right) \Psi_{ \sigma^{\prime}}^{\dagger}\left(\bm{r}^{\prime}, \tau'\right)\right\rangle$,
$F_{\sigma, \sigma^{\prime}}\left(\bm{r}, \bm{r}^{\prime}, \tau-\tau^{\prime}\right)=\left\langle \operatorname{T_{\tau}} \Psi_{\sigma}(\bm{r}, \tau) \Psi_{\sigma^{\prime}}\left(\bm{r}^{\prime}, \tau^{\prime}\right)\right\rangle$, and
$F_{\sigma \sigma^{\prime}}^{\dagger}\left(\bm{r},\bm{r}^{\prime}, \tau-\tau^{\prime}\right)=\left\langle \operatorname{T_{\tau}} \Psi_{\sigma}^{\dagger}(\bm{r}, \tau) \Psi_{\sigma^{\prime}}^{\dagger}\left(\bm{r}^{\prime}, \tau^{\prime}\right)\right\rangle$. Further, passing to the Fourier transform in terms of Matsubara frequencies we obtain 
\begin{align}\label{eq:IIIA11}
&I=-2 e t^{4} \operatorname{Im}\biggl[T \sum_{i \omega_{n}} \sum_{\sigma_{1}, \sigma_{2}} \int_{b 1} \int_{b 2}  d\bm{r}_{b 1} d\bm{r}_{b 1}^{\prime}d\bm{r}_{b 2}d\bm{r}_{b 2}^{\prime}\nonumber \\
&\frac{\partial^{2} F_{-\sigma_{1}, \sigma_{1}}^{\dagger}\left(\bm{r}_{b 1}^{\prime}, \bm{r}_{b 1}, i \omega_{n}\right)}{\partial x_{1}^{\prime} \partial x_{1}} \frac{\partial^{2} G_{\sigma_{1}, \sigma_{2}}\left(\bm{r}_{b 1}, \bm{r}_{b 2}, i \omega_{n}\right)}{\partial x_{1} \partial x_{2}}\nonumber \\
&\frac{\partial^{2} F_{\sigma_{2},-\sigma_{2}}\left(\bm{r}_{b 2}, \bm{r}_{b 2}^{\prime}, i \omega_{n}\right)}{\partial x_{2} \partial x_{2}^{\prime}} \frac{\partial^{2} G_{-\sigma_{1},-\sigma_{2}}\left(\bm{r}_{b 1}^{\prime}, \bm{r}_{b 2}^{\prime},-i \omega_{n}\right)}{\partial x_{1}^{\prime} \partial x_{2}^{\prime}}\biggr].
\end{align}
This expression can be further simplified in our geometry. First, we introduce a more condensed notation for the Green's functions, i.e.,    
$F\left(\bm{r}, \bm{r}^{\prime}, \tau\right)=-F_{\uparrow, \downarrow}\left(\bm{r}, \bm{r}^{\prime}, \tau\right)=F_{\downarrow, \uparrow}\left(\bm{r}^{\prime}, \bm{r},-\tau\right)$, and similarly $F^{\dagger}\left(\bm{r}, \bm{r}^{\prime}, \tau\right)=-F_{\downarrow, \uparrow}^{\dagger}\left(\bm{r},\bm{r}^{\prime}, \tau\right)=F_{\uparrow, \downarrow}^{\dagger}\left(\bm{r}^{\prime}, \bm{r}, \tau\right)$. Second, we take a partial momentum Fourier transform along the direction parallel to the surface of the tunnel boundary. For example, with $\bm{r}_{b}=(x, \bm{r}_{\|})$ and $\bm{r}_{b}^{\prime}=(x^{\prime}, \bm{r}_{\|}^{\prime})$,
\begin{equation}
G(x,x',\bm{k}_\parallel,i\omega_n)=\int d(\bm{r}_\parallel-\bm{r}'_\parallel)e^{-i\bm{k}_\parallel(\bm{r}_\parallel-\bm{r}'_\parallel)}G(\bm{r},\bm{r}',i\omega_n).
\end{equation}
As a result, in the mixed position-momentum representation, Eq. (\ref{eq:IIIA11}) for the current density $j(\phi)=I/S_d$ 
through the interface with the area $S_d$ reduces to
\begin{equation}\label{eq:IIIA12}
\begin{gathered}
j(\phi)=-2 e t^{4} \operatorname{Im}\biggl[T \sum_{i \omega_{n}} \int_{\bm{k}_\parallel} P\left(-\frac{L}{2}, \frac{L}{2}, \boldsymbol{k}_{||}, i \omega_{n}\right)\\
\times\frac{\partial^{2} F\left(\frac{L}{2}, \frac{L}{2}, \boldsymbol{k}_{||}, i \omega_{n}\right)}{\partial x \partial x^{\prime}} 
\frac{\partial^{2} F^{\dagger}\left(-\frac{L}{2},-\frac{L}{2}, \boldsymbol{k}_{||}, i \omega_{n}\right)}
{\partial x \partial x^{\prime}}\biggr],
\end{gathered}
\end{equation}
where the kernel function under the integral is defined by
\begin{equation}\label{eq:IIIA13}
\begin{gathered}
P=\sum_{\sigma}
\left[\frac{\partial^{2} G^+_{\sigma, \sigma}}{\partial x \partial x^{\prime}} \frac{\partial^{2} G^-_{-\sigma,-\sigma}}{\partial x \partial x^{\prime}}-\frac{\partial^{2} G^+_{\sigma,-\sigma}}{\partial x \partial x^{\prime}} \frac{\partial^{2} G^-_{-\sigma, \sigma}}{\partial x \partial x^{\prime}}\right].
\end{gathered}
\end{equation}
Here we used a short-hand notation $G^{\pm}_{\sigma,\sigma'}=G_{\sigma, \sigma'}\left(-\frac{L}{2}, \frac{L}{2}, \pm\boldsymbol{k}_{||}, \pm i \omega_{n}\right)$. We note that a similar formula for the supercurrent appeared in Ref. \cite{Rasmussen2016}. The differences are (i) the model for tunnel barriers, and (ii) effects of SOC were then studied numerically. Equation \eqref{eq:IIIA12} has a transparent diagrammatic representation, which is depicted in Fig. \ref{fig3}. Therefore the problem of finding the current-phase relation and an anomalous phase shift is reduced to deriving $P\left(-\frac{L}{2}, \frac{L}{2}, \boldsymbol{k}_{||}, i \omega_{n}\right)$. This defines our next task, which is the determination of the Green's function in the normal region. 

%$$$$$$$$$$$$$$$$$$$$$$$$$$$$$$$$$$$$$$$$$$$$$$$$$$$$$$$$$$$$$$$$$$$$
%$$$$$$$$$$$$$$$$$$$$$$$$$$$$$$$$$$$$$$$$$$$$$$$$$$$$$$$$$$$$$$$$$$$$

%@@@@@@@@@@@@@
\subsection{Green's functions}
%@@@@@@@@@@@@@

For the planar 2D geometry, with the identification of $k_\parallel\to k_y$, the Green's function in the normal layer satisfies the matrix differential equation,
\begin{equation}\label{eq:IIIA19}
\begin{gathered}
\left(\begin{array}{cc}
i \omega_{n}-\varepsilon_{y}+\frac{1}{2 m} \frac{\partial^{2}}{\partial x^{2}} & -\alpha k_{y}-\alpha \frac{\partial}{\partial x}-h e^{-i \phi} \\
-\alpha k_{y}+\alpha \frac{\partial}{\partial x}-h e^{i \phi} & i \omega_{n}-\varepsilon_{y}+\frac{1}{2 m} \frac{\partial^{2}}{\partial x^{2}}
\end{array}\right)\boldsymbol{\hat{G}}\\=\boldsymbol{\hat{I}}\delta\left(x-x^{\prime}\right),
\end{gathered}
\end{equation}
where $\varepsilon_{y}=k_{y}^{2} / 2 m-\varepsilon_F$, $\boldsymbol{\hat{I}}$ is the $2 \times 2$ identity matrix, and the $2 \times 2$ Green's function matrix $\boldsymbol{\hat{G}}$ in the spin space is given by
\begin{equation}\label{eq:IIIA20}
\boldsymbol{\hat{G}}=\left(\begin{array}{cc}
G_{\uparrow \uparrow}\left(x, x^{\prime}, k_{y}, i \omega_{n}\right) & G_{\uparrow \downarrow}\left(x, x^{\prime}, k_{y}, i \omega_{n}\right) \\
G_{\downarrow \uparrow}\left(x, x^{\prime}, k_{y}, i \omega_{n}\right) & G_{\downarrow \downarrow}\left(x, x^{\prime}, k_{y}, i \omega_{n}\right)
\end{array}\right).
\end{equation}
The rigid boundary conditions are of the form 
\begin{equation}\label{eq:IIIA21}
G_{\sigma \sigma^{\prime}}\left(\pm \frac{L}{2}, x^{\prime}, k_{y}, i \omega_{n}\right)=G_{\sigma \sigma^{\prime}}\left(x, \pm \frac{L}{2}, k_{y}, i \omega_{n}\right)=0.
\end{equation}
For the finite-size system with $-\frac{L}{2}<\{x, x^{\prime}\}<\frac{L}{2}$, the general solution can be written down as a series expansion in eigenfunctions 
\begin{equation}\label{eq:IIIA22}
\begin{gathered}
G\left(x, x^{\prime}, k_{y}, i \omega_{n}\right)\\=\left\{\begin{array}{lll}
\sum_{\lambda, \lambda^{\prime}} C^<_{\lambda, \lambda^{\prime}} e^{\lambda(x+L / 2)-\lambda^{\prime}\left(x^{\prime}-L / 2\right)} & \text { when } & x<x^{\prime} \\
\sum_{\lambda, \lambda^{\prime}} C^>_{\lambda, \lambda^{\prime}} e^{\lambda(x-L / 2)-\lambda^{\prime}\left(x^{\prime}+L / 2\right)} & \text { when } & x>x^{\prime}
\end{array}\right.
\end{gathered}
\end{equation}
Here each term satisfies Eq. (\ref{eq:IIIA19}) at $x \neq x^{\prime}$, when its right-hand side is zero. Therefore, $\lambda$ and $\lambda^{\prime}$ are the solutions of the characteristic equation,
\begin{equation}\label{eq:IIIA23}
\operatorname{det}\boldsymbol{\hat{\Xi}}=0
\end{equation}
with
\begin{equation}\label{eq:IIIA24}
\begin{gathered}
\boldsymbol{\hat{\Xi}}=\\\left(\begin{array}{cc}
\lambda^{2}-2 m\left(\varepsilon_{y}-i \omega_{n}\right) & -2 m \alpha\left(\lambda+k_{y}\right)-2 m h e^{-i \varphi} \\
2 m \alpha\left(\lambda-k_{y}\right)-2 m h e^{i \varphi} & \lambda^{2}-2 m\left(\varepsilon_{y}-i \omega_{n}\right)
\end{array}\right)
\end{gathered}.
\end{equation} 
We first solve this equation exactly at $h=0$. The solutions come in pairs $\pm \lambda_{1}$ and $\pm \lambda_{2}$, where we have chosen $\operatorname{Re} \lambda_{\gamma}>0$ for $\gamma=1,2$:
\begin{equation}\label{eq:IIIA25}
\lambda_{\gamma}^{2}=-\left(k_{0}-\gamma m \alpha\right)^{2}+k_{y}^{2} .
\end{equation}
Here we use the notation $\gamma=1,2$ when $\gamma$ is a subscript and $\gamma=\pm$ when $\gamma$ is a factor (i.e. we use $\gamma=+$ for $\gamma=1$ and $\gamma=-$ for $\gamma=2$). Above we have also defined $k_{0}^{2}=k_{\alpha}^{2}+2 i m \omega_{n} \quad \text {with} \quad \operatorname{Re} k_{0}>0$, $k_{\alpha}^{2}=k_F^2+m^{2} \alpha^{2}$ and $k_{F}=\sqrt{2 m \varepsilon_F}$ being the Fermi momentum in the absence of spin-orbit coupling. We then find linear in magnetic field corrections. Labeling the solutions of Eq. (\ref{eq:IIIA23}) with the notation $\lambda_{\gamma \sigma}$, where $\gamma=1,2$ and $\sigma=\pm$, we obtain
\begin{equation}\label{eq:IIIA26}
\lambda_{\gamma \sigma}=\sigma \lambda_{\gamma} +\gamma \frac{i m h_{y}}{k_{0}} +\gamma \sigma \frac{m k_{y} h_{x}}{\lambda_{\gamma} k_{0}}.
\end{equation}
We further concentrate on the effect of $h_{y}$ exclusively and put $h_{x}=0$. From the example considered within GL approach we saw that the impact of finite $h_x$ is to suppress the critical current.  

Using the eigenvalues $\lambda_{\gamma \sigma}$, the boundary conditions (\ref{eq:IIIA21}), and the constraint imposed on the derivatives of the Green's function at $x=x^{\prime}$, which is obtained by integrating Eq. \eqref{eq:IIIA19} and using the continuity of $\hat{\bm{G}}$, we can solve for the Green's function analytically. For the calculation of the Josephson current, we will need the following derivative of the Green's function matrix in spin space: 
\begin{equation}\label{eq:IIIA28}
\begin{gathered}
\frac{\partial^{2} \hat{\bm{G}}\left(-\frac{L}{2}, \frac{L}{2}, k_{y}, i \omega_{n}\right)}{\partial x \partial x^{\prime}}\\=\frac{4 m k_{0}}{D\left(k_{y}, i \omega_{n}\right)}\left(\begin{array}{cc}
R\left(k_{y}, i \omega_{n}\right) & Q\left(k_{y}, i \omega_{n}\right) \\
-Q\left(-k_{y}, i \omega_{n}\right) & R\left(k_{y}, i \omega_{n}\right)
\end{array}\right).
\end{gathered}
\end{equation}
In the limit $L\gg l_T$ a simplified form of these functions can be taken 
\begin{subequations}
\begin{align}
&D=(\lambda_1\lambda_2-k^2_y-k^2_0+m^2\alpha^2)e^{\beta_{1+}+\beta_{2+}}, \\ 
&R=-\lambda_2(k_0-m\alpha)e^{\beta_{1+}}-\lambda_1(k_0+m\alpha)e^{\beta_{2+}}, \\
&Q=\lambda_2(\lambda_1+k_y)e^{\beta_{1+}}-\lambda_1(\lambda_2+k_y)e^{\beta_{2+}}
\end{align}
\end{subequations}
with the notation $\beta_{\gamma\sigma}=\lambda_{\gamma\sigma}L$. The general analytical expressions for $D\left(k_{y}, i \omega_{n}\right)$, $R\left(k_{y}, i \omega_{n}\right)$ and $Q\left(k_{y}, i \omega_{n}\right)$ are cumbersome and thus not presented here for brevity. 

%@@@@@@@@@@@@@@@@
\subsection{Anomalous phase shift}
%@@@@@@@@@@@@@@@@

We explore Eq. \eqref{eq:IIIA12} in several limiting cases that can be arranged in terms of relations between length or equivalently, energy scales. One tractable limit corresponds to the situation of a weak SOC and a high temperature, when $l_{so}\gg L\gg l_T$. An integration of the analytical expressions in Eq. \eqref{eq:IIIA12} leads to the supercurrent density of the form
\begin{equation}
j(\phi)=j_0\left[\sin(\phi)\cos\left(\frac{2h_yL}{v_F}\right)+\frac{\alpha}{v_F}\cos(\phi)\sin\left(\frac{2h_yL}{v_F}\right)\right],
\end{equation}
which can be rewritten as Eq. \eqref{eq:i-phi-GL} with the phase shift 
\begin{equation}\label{eq:phi0-confined}
\phi_0=\arctan\left[\frac{\alpha}{v_F}\tan\left(\frac{2h_yL}{v_F}\right)\right]\approx\frac{2\alpha h_yL}{v^2_F},  
\end{equation}
provided that $\alpha\ll v_F$ and $h_y\ll v_F/L$. The amplitude of the current scales as $j_0\propto \sqrt{\frac{l_T}{L}}e^{-L/l_T}$, and such an exponential falloff is characteristic for the high-temperature regime \cite{ALO1969}. At lower temperatures, $T\ll v_F/L\ll \Delta$,
we find parametrically the same phase shift, which differs from the expression above only by a numerical factor. However, the temperature dependence of the critical current changes significantly and scales logarithmically with the system size $j_0\propto \ln(v_F/LT)$. This result, however, does not hold in the limit of $T\to0$. The reason is that our oversimplified treatment of the tunneling at SIN boundaries misses the proximity-induced spectral gap $E_g(\phi)$ in the normal region. When temperature becomes smaller than the so-galled minigap, $T<E_g$, the logarithmic dependence is cut off. Furthermore, as the gap depends on the phase across the junction, $E_g\propto \cos(\phi/2)$, this leads to a nonsinusoidal skewed current-phase relation near the gap closure $\phi\sim\pi$, namely, $j_0\propto \sin(\phi)\ln\left[\frac{v_F/L}{E_g(0)}|\cos(\phi/2)|\right]$, see Refs. \cite{Brouwer1997,LKG2006,Whisler2018}. These considerations suggest that unlike the critical current, phase shift has weak temperature dependence, although due to the limitations of the model we can't access the temperature regime below the minigap to make a definitive statement in that limit. 

%$$$$$$$$$$$$$$$$$$$$$$$$$$$$$$$$$$$$$$$$$$$$$$$$$$$$$$$$$$$$$$$$$$$$
\begin{figure}[t!]
\includegraphics[width=0.48\textwidth]{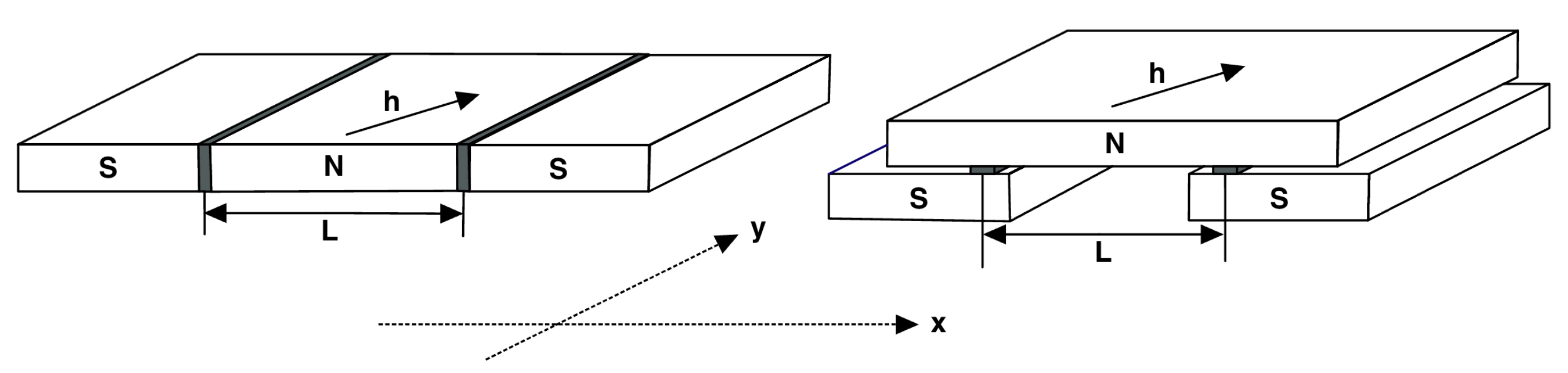} 
\caption{Two complimentary geometries of the planar SINIS devices: the left panel shows confined two-dimensional geometry, and the right panel shows the extended geometry. Tunnel barriers are marked by shaded regions. }  \label{fig5}
\end{figure}
%$$$$$$$$$$$$$$$$$$$$$$$$$$$$$$$$$$$$$$$$$$$$$$$$$$$$$$$$$$$$$$$$$$$$

\begin{figure}[t!]
\includegraphics[width=0.48\textwidth]{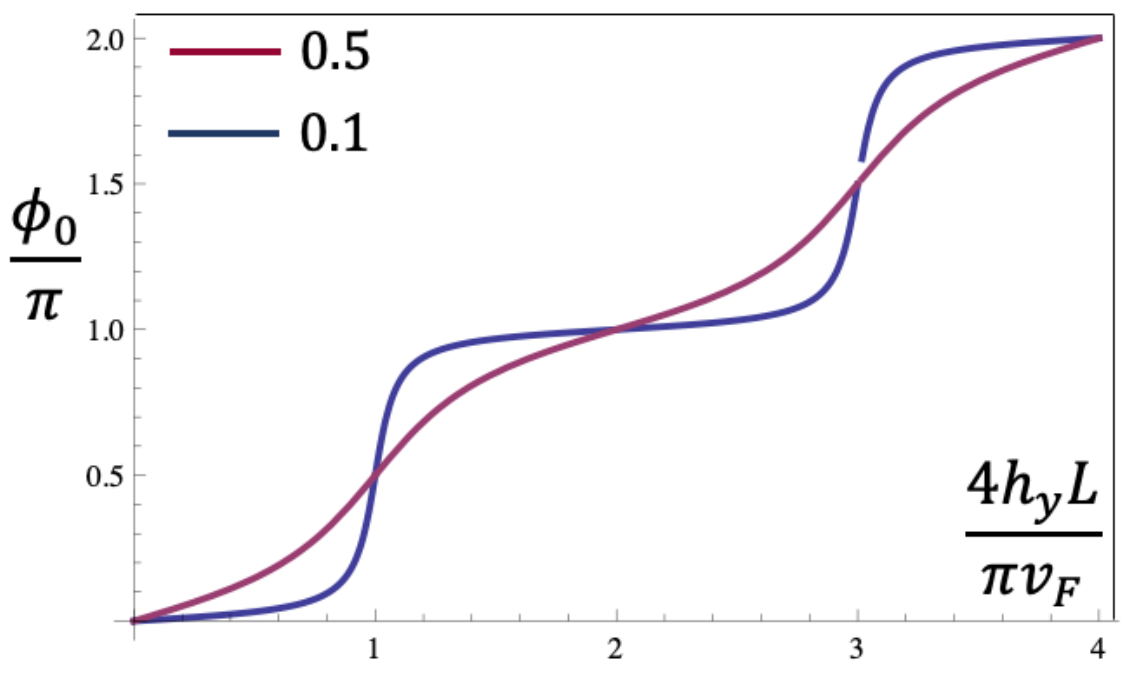} 
\caption{Dependence of the anomalous phase shift $\phi_0$ on the strength of the Zeeman field and the length of the junction plotted for two different ratios of spin-orbit to Fermi velocities $\alpha/v_F=0.1,0.5$.}  \label{fig4}
\end{figure}

Earlier works \cite{Bergeret2015,Konschelle2015} demonstrated that a linear scaling of $\phi_0$ with the strength of the spin-orbit coupling and system size at weak fields is not robust. It was shown that $\phi_0$ is extremely sensitive to the boundary conditions in comparison to the sensitivity to the type of scattering in the bulk of the normal layer. Indeed, for transparent interfaces one finds $\phi_0\propto L^3$. This lead us to consider an alternative version of the model that in-part mimics an extended (rather than rigid) system. For that purpose we took an SINIS junction in which the size of the normal layer in $x$ direction is much larger than the distance $L$ between the two tunnel barriers, i.e., the superconducting leads and the normal parts are in different planes, and the tunneling occurs between those planes, see Fig. \ref{fig5} for the illustration. In this setting, the Green's function is translationally invariant, $G_{\sigma,\sigma'}(x-x',\bm{k}_\parallel,i\omega_n)$, which simplifies analytical calculations. The expression for the current 
can be derived for this geometry, and it takes a form similar to Eq. \eqref{eq:IIIA12}, albeit with the different kernel function under the integral, namely,  
\begin{align}
j(\phi)&\propto
\operatorname{Im}\biggl[T \sum_{i \omega_{n}} \int_{k_y} P\left(-\frac{L}{2}, \frac{L}{2}, k_{y}, i \omega_{n}\right)\nonumber \\ &\times
F\left(\frac{L}{2}, \frac{L}{2}, k_{y}, i \omega_{n}\right)
F^{\dagger}\left(-\frac{L}{2},-\frac{L}{2}, k_{y}, i \omega_{n}\right)
\biggr],
\end{align}
where $P=\operatorname{Tr}\left[\sigma_y\hat{\bm{G}}_+\sigma_y\hat{\bm{G}}^T_-\right]$, with the notation $\hat{\bm{G}}_\pm=\hat{\bm{G}}(-L/2,L/2, \pm k_{y}, \pm i \omega_{n})$ and with the superscript $T$ denoting matrix transposition. For simplicity, we took the tunnel matrix elements to be independent of $k_x$. In this spatially extended model, we derived the following result for the anomalous phase shift in the current-phase relation, 
\begin{align}\label{eq:phi0-extended}
&\phi_0=\frac{h_y}{\varepsilon_F}\left[\frac{L}{l_{so}}+\frac{1}{k^2_Fl_Tl_{so}}\right.\nonumber \\ 
&\left.-\frac{l_{so}}{2L}\left[1-\sqrt[4]{\frac{1}{1+4l^2_T/l^2_{so}}}\cos\left(\frac{2L}{l_{so}}+\psi\right)\right]\right], 
\end{align}
which is applicable in the long junction limit, $L\gg l_T$, and $\psi\simeq l_T/l_{so}$. This expression enables us to consider additional limiting cases. If $l_{so}\ll l_T\ll L$ or $\alpha k_F\gg T$, the first contribution in Eq. \eqref{eq:phi0-extended} is clearly dominant and we again recover the conventional results $\phi_0\sim \alpha h_y L/v^2_F$.  If instead $l_T\ll l_{so}\ll L$, the quartic-root term is of the order $\sim 1$ but the prefactor is still suppressed, and again we find the standard relation $\phi_0\propto L$. Lastly, in the limit $l_T\ll L\ll l_{so}$ the root term is still of order unity but the cosine term should be treated carefully and expanded up to the fourth order to recover the leading behavior, where we find 
\begin{equation}\label{eq:phi0-extended-L3}
\phi_0\approx \frac{2}{3}(k_FL)^2\left(\frac{\alpha}{v_F}\right)^2\left(\frac{\alpha h_yL}{v^2_F}\right).
\end{equation}
Interestingly, this reproduces the behavior known from the context of ballistic junctions with transparent interfaces \cite{Konschelle2015}. To generalize these results beyond the perturbation theory in the Zeeman field, and in a broader range of parameters, one has to rely on the numerical solution. In Fig. \ref{fig4} we sketch the characteristic dependence of the anomalous phase shift at higher fields and at different spin-orbit to Fermi velocity ratios.

%######################################################################
%######################################################################
%######################################################################
%######################################################################
%######################################################################

\section{Summary and discussion}\label{sec:summary}

In this work we have studied planar SINIS Josephson junctions using both phenomenological Ginzburg-Landau and microscopic Green's function formalisms in the clean limit. To describe the normal layer, we took the model of a two-dimensional electron gas with the Rashba spin-orbit coupling and included an in-plane Zeeman field directed at an arbitrary angle with respect to the SN interfaces. Two complimentary device geometries were analyzed as depicted in Fig. \ref{fig5}. We have shown that the supercurrent-phase relation in these systems acquires an additional anomalous phase shift $\phi_0$ whose magnitude can be continuously tuned by the field component directed along the interface, whereas the component along the junction modulates the magnitude of the current. The main results of this work are expressions of the current density derived in the GL formalism Eq. \eqref{eq:GL12}, a formula for the Josephson current in the Rashba model Eq. \eqref{eq:IIIA12}, and extracted asymptotic expressions of the anomalous phase shifts Eqs. \eqref{eq:phi0-confined} and \eqref{eq:phi0-extended} that describe the crossover regimes in the field strength and system size as compared to other relevant length scales in the problem.  

As our calculations are limited to the clean limit, we find it useful to contrast our findings with the complementary results obtained in the opposite disordered limit. This gives a broader perspective on the problem and helps us to place our study in the context of existing earlier works. We limit such comparative analysis only to the case of the Rashba model of SINIS devices as other systems may introduce additional features not discussed in our work. From Ref. \cite{Bergeret2015} we deduce that in the disordered limit and at temperatures $T>\Delta(T)$, the anomalous phase shift takes the form 
\begin{equation}\label{eq:pih0-diffusive}
\phi_0\approx\frac{\kappa^2_h}{2\kappa^2_T}\tanh(\kappa_\alpha L)\left[\frac{\kappa_T L}{\tanh(\kappa_T L)}\pm1\right].
\end{equation}
Here $\kappa_h=\sqrt{2h/D}$, $\kappa_T=\sqrt{2\pi T/D}$, $\kappa_\alpha=\tau\alpha^3m^2/4$, with $\tau$ being the elastic scattering time on the quenched short-range disorder potential and $D=v^2_F\tau/2$ being the corresponding diffusion constant. The plus/minus sign in Eq. \eqref{eq:pih0-diffusive} describes different boundary conditions: plus stands for tunnel barriers while minus for transparent interfaces. Equation \eqref{eq:pih0-diffusive} shows that for long junctions, $\kappa_T L\gg 1$, the result for $\phi_0$ is the same in both cases. This limit is simultaneously compatible with the condition $\kappa_\alpha L\ll1$, and therefore the phase shift has an unusual length dependence, $\phi_0\sim (\alpha hL/v^2_F)\sqrt{l_T/l}(Ll/l^2_{so})\propto L^2$, where $l=v_F\tau$ is the elastic mean free path. In contrast, for short junctions, $\kappa_TL\ll1$, the situation is different. In the case of tunnel barriers, one finds $\phi_0\sim (\alpha hL/v^2_F)(ll_T/l^2_{so})\propto L$. Instead, for transparent interfaces, one recovers the same result as in the clean limit Eq. \eqref{eq:phi0-extended-L3}, as parameters of the disorder surprisingly cancel out. The physical picture behind this coincidental result is not immediately obvious. The result only suggests that in short junctions boundary conditions play a decisive role, and a finite barrier resistance of the SN contacts enhances the value of $\phi_0$.

\section*{Acknowledgments}

We thank E. Rossi and J. Shabani for the communication regarding the SQUID magnetometry measurements of the anomalous phase shifts reported in Ref. \cite{Shabani2020}. The financial support for this work at the University of Wisconsin-Madison was provided by the National Science Foundation, Quantum Leap Challenge Institute for Hybrid Quantum Architectures and Networks, Grant No. 2016136. We acknowledge support from the ANR through Grants No. ANR-17-PIRE-0001 and No. ANR-21-CE30-0035 (M.H. and J.S.M). 

\bibliography{biblio}
\end{document}